# Sécurité des systèmes critiques et cybercriminalité : vers une sécurité globale ?

Walter SCHÖN
*Laboratoire Heudiasyc UMR CNRS 6599,*
*Université de Technologie de Compiègne,*
*B.P. 20529, F-60205, Compiègne - France*

**Résumé**
Cet article se propose de mettre en perspective pour les systèmes informatiques dits « critiques » les problématiques de résistance à la cybercriminalité (dont l'origine est donc une action humaine avec intention de nuire), et de faculté à éviter des comportements catastrophiques suite à des événements d'autre origine, pouvant être des défaillances internes des composants matériels, des perturbations de l'environnement, voire des fautes humaines involontaires dans la conception ou l'exploitation du système. De manière assez fâcheuse, en langue Française les deux problématiques sont désignées par le même mot « sécurité », traduisant de fait les deux mots Anglais security et safety, ce qui est souvent source de confusion y compris dans l'esprit d'un public averti. Employer à la place le mot « sûreté » souvent utilisé également dans le cadre de la lutte contre la criminalité (mais utilisé également dans un sens totalement différent dans d'autres communautés dont celle du nucléaire), ne fait qu'augmenter cette confusion fréquente qui au demeurant n'est pas totalement due au hasard : d'une part les deux termes ont la même étymologie (du latin securitas) et d'autre part comme on le développe dans l'article les techniques mises en œuvre ont un lien de parenté même s'il n'est pas toujours apparent au premier regard. La sécurité de systèmes informatiques modernes communicants comme on peut en trouver par exemple dans les systèmes de signalisation ferroviaire récents, rend d'ailleurs incontournable la maîtrise conjointe de la « safety », (parfois appelée sécurité innocuité) et de la « security » (parfois appelée sécurité confidentialité) pour atteindre une « sécurité globale ».
*Mots clés : Sécurité Innocuité, Sécurité Confidentialité, Cybercriminalité, Systèmes Critiques.*

# 1 Introduction

Cet article aborde la problématique des systèmes informatiques dits « critiques » à savoir dont un dysfonctionnement quel qu'en soit l'origine peut avoir des conséquences graves en termes de dommages aux personnes, aux biens et à l'environnement. La problématique de la cyber délinquance, de la cybercriminalité voire du cyber-terrorisme ne peut en effet plus se limiter aux intrusions passives ou actives à des informations utilisées par la suite pour mener des actions de terrorisme « classique » (par exemple). A l'ère où nos avions, trains, voitures, centrales de production d'énergie… sont contrôlés et commandés par des systèmes informatiques qui communiquent de plus en plus, il faut impérativement avoir réfléchi à la problématique de l'intrusion malveillante dans ces systèmes dans des buts criminels ou terroristes. Il doit par conséquent être plus difficile de provoquer une catastrophe technologique par des moyens informatiques que par les moyens « classiques ». Cet article se propose de montrer que les techniques de couverture des dysfonctionnements « accidentels », classiquement mises en œuvre pour les systèmes critiques, ne sont pas si éloignées de celles permettant la couverture des dysfonctionnements dus à la malveillance humaine. La possibilité de ces dernières peut amener simplement à mener certaines techniques avec une rigueur accrue (la cyber malveillance étant donc dans ce cas également couverte), dans certains cas une extension est nécessaire car la technique ne couvre à la base que les événements « accidentels ».

L'article est donc organisé comme suit :

Dans une première section on définit les principaux concepts et on donne la typologie maintenant classique de ce qu'on appelle les fautes, erreurs et défaillances des systèmes informatiques.

Dans une seconde section on donne les principales techniques de couverture des fautes « accidentelles » techniques, d'environnement ou humaines sans intention de nuire (ce que l'on appelle communément les « erreurs humaines », mais qui peuvent affecter outre l'exploitant le concepteur ou le mainteneur).

Dans la troisième et dernière section, on examine dans quelle mesure les fautes humaines à caractère volontairement nuisible (quel que soit le rôle de l'individu malveillant vis à vis du système : il faut envisager qu'il puisse être l'un des concepteurs, l'un des exploitants ou des mainteneurs, ou un simple utilisateur ou extérieur au système) sont où non couvertes par les techniques vues dans la seconde section et on identifie les points à renforcer.

## 2 Des fautes, des erreurs et des défaillances… incluant l'intention de nuire

Selon une terminologie maintenant normalisée, les comportements indésirables pouvant affecter les systèmes industriels (y compris les systèmes informatiques) sont désignés par le vocable *défaillance*, que le dictionnaire définit comme l'arrêt du fonctionnement *normal*, et les normes internationales applicables au domaine comme la cessation de l'aptitude à effectuer une fonction *requise*. Il convient donc de souligner à ce stade que l'étude des défaillances commence par conséquent par définir où finit le normal et où commence l'anormal, ou ce qui revient au même quelles sont les fonctions requises et celles qui ne le sont pas. L'approche classique du problème implique donc de partitionner les états du système en deux classes mutuellement exclusives où le fonctionnement du système est décrété normal (respectivement anormal), quitte à définir au besoin plusieurs partitions (et donc plusieurs points de vue pour la notion de défaillance d'un même système).

La science pour l'ingénieur qui traite de manière générale des défaillances (et naturellement des moyens et méthodes à mettre en œuvre pour les maîtriser) est appelée *sûreté de fonctionnement* [5]. Elle regroupe les quatre disciplines que sont la Fiabilité (mesurée par une probabilité pour un système de fonctionner sans défaillance sur une durée donnée), la Disponibilité (mesurée par la probabilité instantanée pour un système d'être non défaillant), la Maintenabilité (mesurée par une probabilité de parvenir à réparer un système défaillant sur une durée donnée), et la *Sécurité*, l'acronyme FDMS étant synonyme de sûreté de fonctionnement. La sécurité (sujet du présent article) peut se définir de manière pour l'instant non formalisée comme l'aptitude pour un système d'éviter de provoquer des événements dits catastrophiques (en termes de dommages aux personnes ou aux biens). Elle regroupe de fait deux notions différentes (bien que liées pour les systèmes informatiques ainsi qu'on le développe par la suite) que sont la robustesse (vis à vis de possibles conséquences catastrophiques) en présence de :

- Événements internes ou externes au système qui ne sont pas des actions humaines avec intention de nuire.

- Événements internes ou externes au système qui sont la conséquence d'actions humaines avec intention de nuire (la cybercriminalité sujet du présent numéro de la revue en étant la déclinaison pour les systèmes informatiques).

La terminologie Anglaise, plus précise dans ce cas distingue ces deux notions par deux termes différents, à savoir safety pour la première (parfois traduit en Français avec la précision sécurité *innocuité*) et security pour la seconde[1] (parfois traduit en Français avec la précision sécurité confidentialité, la confidentialité étant l'une des problématiques essentielles, bien que pas la seule, de la security au moins en ce qui concerne les systèmes informatiques).

D'une manière générale donc, si l'on tient à être précis, le terme sûreté (tout court) est à éviter car source de confusion du fait d'une signification différente suivant les domaines dans lequel il est employé. On sait maintenant que la sûreté *de fonctionnement* inclut les aspects FDMS, le mot sûreté ayant une signification bien différente suivant que l'on parle de sûreté du territoire ou des espaces ouverts au public, ou bien de la sûreté nucléaire.

Pour les systèmes informatiques, l'approche qui fait maintenant référence [8] distingue sous le vocable d'entraves à la sûreté de fonctionnement, les trois niveaux que sont les *fautes*, les *erreurs* et les *défaillances*. A l'origine de la défaillance (définie plus haut) est donc l'erreur (état erroné du système, par exemple un bit erroné dans un mot mémoire) qui est elle même la conséquence d'une faute (par exemple un dysfonctionnement interne du circuit intégré support physique de cette mémoire, ce qui à l'échelle du composant est une défaillance). La théorie de la sûreté de fonctionnement des systèmes informatiques se développe donc autour de ces chaînes causales : fautes => erreurs => défaillances (parfois hautement ramifiées du fait du phénomène appelé *propagation des erreurs*). L'importance de la distinction entre faute et erreur est liée au fait qu'une faute peut rester latente ou dormante (par exemple une faute dans un mot mémoire correspondant à une instruction de programme le reste tant que l'instruction n'est pas exécutée), et ne se traduire par une

---

[1] La FDMS se traduisant donc RAMSS comme Reliability Availability Maintainability Safety & Security

erreur qu'une fois qu'elle est *activée*. La typologie des causes de défaillance des systèmes informatiques s'obtient donc d'une manière assez claire à partir d'une classification des fautes qui en sont l'origine première suivant cinq critères :

- La cause : physique ou humaine
- La nature : accidentelle, intentionnelle avec intention de nuire, intentionnelle sans intention de nuire
- La phase de création de la faute : en développement, en opération
- La situation par rapport au système : interne ou externe
- La persistance temporelle : permanente ou transitoire

Sur les théoriquement 48 types différents de fautes auxquels conduirait cette classification, 8 seulement apparaissent pertinents, répartis en trois grandes catégories :

- Les fautes physiques accidentelles
    - Internes en développement (défauts de fabrication)
    - Internes en opération (défaillances de composant du support matériel)
    - Externe en opération (perturbations de l'environnement, par exemple rayonnements)
- Les fautes humaines accidentelles ou intentionnelles sans intention de nuire
    - Internes en développement (traditionnellement appelées « bogues »)
    - Externes en opération (fautes de l'opérateur exploitant souvent appelées « erreurs humaines »)
- Les fautes humaines avec intention de nuire
    - Internes en développement (fautes internes introduites intentionnellement par le développeur en général pour constituer une faiblesse exploitée ultérieurement en vie opérationnelle type porte dérobée ou bombe logique)
    - Internes en opération (à cette rubrique on classe les virus et vers qui doivent prendre place en interne bien que l'origine de la contamination soit nécessairement externe)
    - Externes en opération (à cette rubrique on classe les intrusions de tous autres type profitant des ouvertures du système sur l'extérieur)

La sécurité innocuité va donc être l'aptitude du système à éviter les événements catastrophiques consécutivement aux fautes des deux premières catégories (y compris les fautes humaines accidentelles, le développement d'un système « safe » implique des méthodes permettant de maîtriser les « bogues » et une analyse des risques d'erreur humaine en opération). La sécurité confidentialité va quant à elle être l'aptitude à éviter de provoquer des événements catastrophiques consécutivement aux fautes de la troisième catégorie.

L'introduction sans cesse croissante de systèmes informatisés dans des domaines où la préoccupation de « safety » est essentielle (tels le transport, y compris automobile), que l'on appellera désormais *systèmes critiques* dans la suite de cet article, et surtout le fait qu'un nombre croissant de ces systèmes utilisent des communications ouvertes (par transmission radio) implique de prendre également en compte la « security » pour parvenir à une maîtrise de la « sécurité globale ». Ainsi les transports aérien et ferroviaire, qui sont entrés par l'introduction de systèmes de contrôle-commande numériques communicants dans l'ère de la « cyber-safety » doivent aussi prendre en compte la « cyber-security » pour éviter, car étant d'ailleurs la cible privilégiée des terroristes, qu'ils deviennent celle des cyber-terroristes[2].

## 3 Techniques de couverture des fautes autres que celles liées à la malveillance humaine

Dans cette section, on reprend la classification des fautes établie dans la section précédente et on décrit les principales techniques mises en œuvre pour faire en sorte que les fautes autres que celles liées à la malveillance humaine ne conduisent à des défaillances catastrophiques. On passe ainsi en revue les principales techniques de sécurité innocuité avant de voir dans la section suivante dans quelles mesures celles-ci couvrent également, ou peuvent être étendues pour couvrir également, les tentatives de malveillance humaine.

---

[2] Il doit ainsi être par exemple plus difficile de provoquer une catastrophe ferroviaire par intrusion malveillante dans les communications sol / bord des systèmes de signalisation modernes, que par des méthodes de terrorisme « à l'ancienne » impliquant l'intrusion physique sur les lieux, donc plus risquées du point de vue du terroriste.

### 3.1 Fautes physiques accidentelles internes en développement

Les fautes physiques accidentelles internes en développement (défauts de fabrication) sont en tout premier lieu couvertes par la maîtrise de la qualité en fabrication. Ce domaine est un sujet à part entière qui n'est pas celui de cet article. Notons toutefois en complément qu'une telle faute qui aurait franchi cette barrière peut rester dormante jusqu'à son activation en opération (où elle provoque une erreur). Certaines techniques développées ci-après, conçues pour couvrir d'autres types de fautes le font en détectant les erreurs induites (et par conséquent peuvent également couvrir certains défauts de fabrication).

### 3.2 Fautes physiques accidentelles internes ou externes en opération

La couverture des fautes physiques accidentelles internes ou externes en opération sont l'un des sujets majeurs de la sécurité innocuité, à savoir les moyens à mettre en œuvre pour prévenir tout comportement pouvant causer des dommages de gravité importante d'un système correctement conçu et fabriqué, mais subissant en opération des défaillances internes de ses composants électroniques, ou des perturbations du fonctionnement de ses composants par l'environnement. Toutes les techniques de couverture de ces fautes reposent sur la notion de *redondance*.

#### 3.2.1 Redondance matérielle

La plus connue de ces techniques est la *redondance matérielle* qui consiste à mettre en place plusieurs unités effectuant les mêmes traitements, la sécurité reposant sur l'accord sur les résultats (unanimité ou vote majoritaire). C'est ainsi que sont conçus les systèmes informatiques embarqués qui effectuent les commandes de vol des avions actuels ou des engins spatiaux. L'hypothèse sous-jacente est dans ce cas *l'indépendance* des fautes accidentelles pouvant affecter les unités redondantes, tout *mode commun* (causes commune pouvant en affecter plusieurs voire toutes), mettant à mal l'efficacité de la redondance. Ce point doit être soigneusement réfléchi (en particulier pour ce qui concerne le partage d'alimentation, de données d'entrées ou de sorties communes, voire de conditions d'environnement semblables) afin d'atteindre effectivement le niveau de sécurité théorique. Cela amène par exemple pour certaines architectures embarquées dans les domaines du transport à redonder les bus d'entrée et de sortie ou à disséminer les unités redondantes sur différentes parties du véhicule (diversification géographique). Lorsque les unités redondantes sont de même technologie, on parle de *redondance homogène*. Dans le cas où la diversification technologique a été volontairement recherchée, on parle de *redondance hétérogène*. L'idée sous-jacente est dans ce cas de couvrir également de possibles fautes humaines internes en développement ayant pu affecter les briques technologiques utilisées (à savoir essentiellement les processeurs). On sait ainsi que les premières versions des processeurs Pentium souffraient de fautes de conception qui affectaient certaines opérations, qui pouvaient donc être détectées par diversification technologique (la même faute affectant un processeur de fournisseur différent étant extrêmement improbable). Notons donc au passage que, conçue à la base pour couvrir les fautes de conception accidentelles ou sans intention de nuire, la technique peut couvrir une faute malveillante lors de la conception du processeur, théoriquement envisageable bien que difficilement réalisable en pratique et d'un usage moins aisé que l'insertion de code malveillant dans un logiciel applicatif, qui est par contre l'une des problématiques majeures de la cybercriminalité. La couverture des fautes de conception des logiciels donne parfois lieu à une technique comparable à savoir la redondance de développement (plusieurs versions du logiciel étant développées sur la base des mêmes spécifications par des équipes indépendantes). Outre son coût élevé qui la rend rarement mise en œuvre en pratique, la technique a l'inconvénient de laisser ouverte la question de possibles modes communs (les solutions de conception retenues par les équipes indépendantes pouvant s'avérer voisines pour de simples raisons culturelles), et de ne pas couvrir les fautes de la spécification initiale. Pour ces raisons, bien que couvrant théoriquement la cybercriminalité en provenance d'éléments infiltrés dans les équipes de développement (à condition bien sûr que les recrutements au sein de ces équipes soient également suffisamment indépendants), problématique importante et probablement trop peu traitée à ce jour, cette technique ne nous semble pas à recommander.

**3.2.2 Redondance informationnelle**
Une autre technique consiste à redonder non pas l'unité de traitement, support physique compris, mais *l'information* manipulée par l'unité de traitement, afin de permettre un contrôle en ligne et en temps réel de la

*vraisemblance* de l'information manipulée et des opérations effectuées. On entre ici dans le domaine des codes détecteurs et correcteurs d'erreurs, technique majeure de la sûreté de fonctionnement des systèmes informatiques. D'un principe très simple, consistant à traiter outre l'information principale, une information en partie redondante permettant le contrôle, sa mise en œuvre peut couvrir un très vaste champ, qui va du bit de parité (technique de base utilisée dès les premiers échanges de données informatiques par modem) couvrant une faute simple sur un message, à des techniques de codage beaucoup plus élaborées permettant de détecter avec un niveau de probabilité très élevé toute altération d'une donnée, voire de permettre de la corriger. Le but de cet article n'est pas de détailler les différentes techniques de codages pouvant être mises en pratique, dans la suite sont donc brièvement décrites les principales pour les systèmes informatiques critiques. Les codes les plus utilisés permettant de vérifier l'intégrité d'une donnée par exemple à l'arrivée d'un message sur un réseau sont les codes type CRC (codes à redondance cyclique basés sur des opérations polynomiales) qui suivant le polynôme utilisé permettent avec un taux de confiance plus ou moins élevé de garantir que la donnée n'a pas été modifiée accidentellement. Certains codes type codes de Hamming permettent même de corriger des fautes accidentelles (en ramenant la donnée détectée erronée car hors code, à la donnée dans le code la plus proche ce qui implique d'ailleurs des hypothèses sous-jacentes sur le nombre de fautes ayant pu affecter le message). Enfin certaines techniques dont le monoprocesseur codé [9, 10]], mis en œuvre avec succès dans le ferroviaire[3] utilisent un code arithmétique pour contrôler la vraisemblance des opérations. Pour prendre une image simple mais assez représentative, le processeur codé procède comme l'écolier vérifiant la vraisemblance de sa multiplication par l'ancestrale technique de la « preuve par 9 ». Outre les opérandes, une information en partie redondante (le modulo 9 des opérandes) permet de vérifier que l'opération est « vraisemblable ». Certes le processeur utilise un modulo plus efficace que 9 (nombre choisi pour l'algorithme écolier pour de simples raisons de faisabilité pratique des calculs à la main) : à savoir un grand entier dit « clé du code » sur 32 ou 48 bits qui rend l'erreur non détectée très improbable (probabilité de l'ordre de 1/A), mais l'idée sous-jacente est bien la même. Le processeur codé manipule des données codées constituées de deux parties : la partie dite fonctionnelle qui contient la valeur de la variable, et une partie dite code qui contient une information en partie redondante permettant de vérifier la vraisemblance des opérations. Ces opérations doivent par conséquent être adaptées : les opérations arithmétiques sont remplacées par les Opérations Elémentaires (OPELs) manipulant les deux parties des données codées, en effectuant l'opération arithmétique dans la partie fonctionnelle, et une opération permettant de conserver la cohérence du codage dans la partie code. La mise en pratique nécessite toutefois quelques compléments, afin de garantir que toutes les erreurs vraisemblables d'un processeur soient détectées (c'est d'ailleurs l'une des faiblesses de la preuve par 9 de ne pas détecter certaines erreurs naturelles des calculs à la main). Or prendre une variable pour une autre est une simple erreur d'adresse qui n'est détectée qu'au prix de l'ajout dans le champ code d'une *signature statique* permettant d'identifier les variables. Cette signature est conçue pour être *pré-déterminable* hors ligne (plusieurs exécutions d'une même instruction vont pouvoir conduire à des valeurs différentes des variables, mais les signatures seront toujours les mêmes). Les signatures attendues peuvent donc être embarquées dans des PROMs et toute confusion entre opérandes pourra ainsi être détectée (avec une probabilité très élevée car les signatures peuvent prendre toutes les valeurs entre 0 et la clé du code A, les « collisions de signature » sont donc très improbables). De même afin de couvrir tout problème de « fraîcheur » des données (ces applications embarquées temps réel étant toujours exécutées de manière cyclique, le champ code porte également une date. A noter que la prédétermination des signatures des variables (nécessairement fonction des opérations qui ont permis de les calculer, et des signatures des variables opérandes de ces opérations, les variables de base ayant des signatures tirées au hasard) est effectué par un outil appelé Outil de Prédétermination des Signatures qui utilise le code source (et agit donc selon un processus parallèle et analogue à celui de la compilation). Toute faute quelle qu'en soit l'origine (y compris « bogue » dans le compilateur) ayant affecté le processus de compilation est donc détectée, sauf cas très improbable de faute affectant de manière parallèle et analogue le processus de prédétermination des signatures.

---

[3] Utilisent en particulier cette technique les systèmes de signalisation dits TVM sur TGV, le système SACEM [11] sur la ligne A du RER Parisien ainsi que le système d'automatisation de l'exploitation des trains de la ligne 14 du métro de Paris [12].

**3.3 Fautes humaines accidentelles internes en développement**

Les fautes humaines accidentelles internes en développement ou « bogues » sont couvertes par les méthodologies de développement adaptées au niveau de criticité de l'application envisagée. Ainsi pour prendre l'exemple du ferroviaire, la norme de référence (EN 50128 [2]) définit comme obligatoire, hautement recommandé ou recommandé l'usages de certaines techniques de développement pouvant aller jusqu'au développement totalement formel avec le système de preuves associé, selon le niveau d'intégrité de la sécurité (Safety Integrity Level : SIL) envisagé, niveau pouvant aller de 0 (aucune exigence particulière de sécurité) à 4 (logiciel pouvant entrainer des conséquences catastrophiques sur les personnes et les biens). C'est pour couvrir ce type de fautes que sont prévues les activités de Vérification et Validation qui peuvent être très variées (tests, simulation exhaustive de modèles, preuves…). Pour les systèmes critiques, il est de toute première importance que ces activités soient réalisées par une équipe totalement indépendante de l'équipe de développement (ce qui n'empêche toutefois pas cette dernière d'effectuer ses propres tests). Nous y reviendrons.

**3.4 Fautes humaines accidentelles externes en opération**

Les fautes humaines accidentelles externes en opération sont le domaine privilégié des études d'interface homme / système technologique et touchent là aux aspects ergonomie et psychologie cognitive. Ces aspects, de toute première importance pour les systèmes critiques dont la plupart des incidents ou accidents majeurs de fonctionnement sont liés à ce que l'on appelle communément « l'erreur humaine », ne sont pas le cadre de cet article qui se propose de montrer que certaines techniques de robustesse aux fautes accidentelles, couvrent de fait ou peuvent être étendues pour couvrir, les malveillances humaines.

## 4 Techniques de couverture des fautes liées à la malveillance humaine (cyber délinquance / criminalité ou terrorisme)

Ayant dans ce qui précède décrit les principales techniques de couverture des fautes accidentelles quelles qu'en soit l'origine (physique ou humaine donc) on décrit donc dans la suite comment ces techniques ou une extension de ces techniques permet de couvrir certaines fautes humaines avec intention de nuire, problématique de la cyber délinquance, cyber criminalité ou cyber terrorisme.

### 4.1 Fautes malveillantes internes en développement

Le problème des fautes malveillantes internes en développement est de toute première importance, probablement sous-estimée à ce jour. Un fragment de code malveillant embarqué dans une application critique par l'un de ses développeurs malintentionné quelles que soient ses motivations (agissant comme « taupe » pour le compte d'une organisation terroriste, d'une puissance étrangère, ou simplement mécontent de son augmentation de salaire de fin d'année) représente en effet une vulnérabilité importante pour les systèmes critiques dont la couverture n'est guère évidente. De plus après avoir été embarquées en secret par les individus malveillants infiltrés dans les équipes de développement, ces « bombes logiques » comme on les appelle peuvent rester dormantes dans l'attente d'un signal secret pouvant être envoyé ultérieurement (pour tout système communicant avec l'extérieur d'une manière ou d'une autre, pour lequel la bombe logique est une porte dérobée secrète). Pour l'anecdote, on rappellera que bien avant d'occuper les fonctions de ministre de l'Economie, des Finances et de l'Industrie[4], Thierry Breton avait décrit dans un thriller technologique à succès[5] les mécanismes par lesquels une telle bombe logique pouvait être utilisée comme une arme redoutable en période de guerre froide, en permettant de commander la défaillance des systèmes informatiques d'un ennemi grâce à un signal secret dissimulé dans des données que ces systèmes consultaient régulièrement sur le réseau (en l'espèce il s'agissait de valeurs de données météorologiques bien précises sur des iles bien précises). Parmi toutes les techniques de robustesse aux fautes décrites plus haut, très peu permettent de couvrir ce type de cyber-malveillance interne aux équipes de développement. La redondance de développement le permet en partie (sous réserve de l'indépendance suffisante des équipes et de leur recrutement permettant avec une confiance raisonnable d'exclure leur infiltration malveillante par des éléments coordonnés), mais on en a déjà signalé le coût et la difficulté de mise en œuvre. Dans les faits la couverture la plus efficace est liée aux activités de vérification et validation qui sous réserve d'une couverture exhaustive de toutes les branches du code (critère souvent utilisé pour les tests dits « boite blanche ») vont donc passer par les fragments de code de la bombe logique et en détecter les fonctionnalités malveillantes. On voit ici clairement la nécessité absolue de faire réaliser les activités de vérification et validation par une équipe indépendante de l'équipe de développement, indépendante en termes de recrutement, pour exclure toute action coordonnée de l'une introduisant la bombe logique et de l'autre (complicité par non détection volontaire), indépendante également dans le processus d'élaboration des cahiers de tests et dans les documents ayant permis cette élaboration[6].

---

[4] De 2005 à 2007 dans les gouvernements de Jean-Pierre Raffarin puis Dominique de Villepin

[5] Sofwar « la guerre douce » titre qui est évidemment un jeu sur le mot Anglais « Software » qui désigne le logiciel.

**4.2 Fautes malveillantes internes en opération**

Les fautes internes humaines malveillantes en opération (virus, vers, chevaux de Troie etc.) concernent pour l'instant peu, fort heureusement, les systèmes critiques car moins ouverts sur l'extérieur que ne le sont les ordinateurs personnels (pratiquement tous reliés à Internet à l'heure actuelle) et du fait que les opérations de maintenance (type mises à jour logicielles, installations…) y sont encadrées par des procédures rigoureuses mises en œuvre par du personnel qualifié. La propagation d'un virus, ou d'un ver sur des réseaux de machines personnelles est le plus souvent le fait d'applications de provenance douteuse (téléchargées ou reçues par mail), que l'utilisateur exécute voire installe (quand il en a les droits) néanmoins par simple curiosité ou pour leur aspect plaisant ou ludique supposé. Une telle vulnérabilité, qui a pour conséquence la florissante industrie des antivirus sur les machines personnelles, n'existe heureusement pas pour les systèmes critiques qui n'ont à exécuter que les applications pour lesquelles ils ont été conçus et dont la maintenance est soigneusement organisée. Ce dernier point apparaît d'ailleurs sous cet éclairage comme crucial car la cyber-délinquance par personne malveillante infiltrée dans le personnel de maintenance, est donc après la cyber-délinquance en développement évoquée dans la sous-section précédente, une deuxième voie d'entrée aux actions malveillantes concernant les systèmes critiques. Ce point est également probablement un point sensible à l'époque où nos automobiles se voient appliquer régulièrement des « patchs » logiciels concernant une partie ou une autre de l'informatique embarquée. Ce type de maintenance doit demeurer réalisable par le seul personnel habilité faute de quoi on risque de voir se répandre des automobiles aux performances « bricolées » (ce qui existe visiblement déjà de manière assez marginale) et également d'ouvrir la porte aux virus embarqués sur automobiles… dont on imagine la gravité.

---

[6] On touche là probablement une limite car les deux équipes finissent toujours par partager certains documents dont les spécifications fonctionnelles qui ne sont pas faites en double sauf « redondance de développement » dont on a déjà signalé les inconvénients.

**4.3 Fautes malveillantes externes en opération**

Les fautes malveillantes externes en opération (intrusion de tous types exploitant les communications du système avec l'extérieur) constituent l'essentiel de la problématique de la cyber-délinquance / criminalité ou terrorisme. C'est le cœur des activités des pirates ou « hackers » de tous types dont les motivations sont d'ailleurs très diverses, allant du simple jeu ou défi (pour accéder à des données protégées ou pour modifier un site web), jusqu'à de véritables actions criminelles ou terroristes organisées, en passant par des actions frauduleuses de diverses natures (détournements de moyens de paiement, téléchargements illégaux etc.).

Moins ouverts sur l'extérieur, les systèmes critiques sont à ce jour moins concernés par ces phénomènes, mais la situation est en pleine évolution. En effet de plus en plus de systèmes de ce type, en particulier pour le contrôle-commande des systèmes de transport utilisent des communications numériques (en particulier entre bord et sol ou entre plusieurs mobiles, dans le cas du transport). Ces communications se font de plus en plus par liaison radio en exploitant les réseaux type téléphonie mobile (ou une déclinaison particulière dédiée comme le GSM-R pour les applications de signalisation ferroviaire normalisée Européenne ERTMS), ou des protocoles normalisés (il est probable de voir apparaître sous peu des applications au domaine du transport basées sur des communications wifi). Il est donc clair que c'est dans ces communications que se situe potentiellement le point faible objet de possibles attaques de cyber-délinquants. Le développement de tels systèmes communicants (il est symptomatique à cet égard que les systèmes modernes de contrôle des trains en applications urbaines soient appelés « CBTC » *Communication Based Train Control*) ne fera qu'accentuer cette tendance qu'il convient de traiter avec le plus grand sérieux. Ces cyber-attaques sur systèmes de transport ne sont d'ailleurs déjà plus de la science fiction comme en témoigne l'incident sérieux survenu en Janvier 2008 dans la ville de Lodz en Pologne où un jeune Polonais de 14 ans est parvenu à faire dérailler un tramway [6] en s'introduisant frauduleusement dans une communication bord-sol par infrarouges ayant pour fonction de télécommander les aiguilles depuis la cabine de conduite du train. Certes le garçon avait eu la possibilité de s'introduire préalablement frauduleusement dans le dépôt du tramway pour y dérober des objets et documents lui ayant facilité la tâche : la première faille de « security » fut donc, comme le plus souvent, un accès trop facile aux locaux ou aux personnes responsables de ces locaux. La technique utilisée par la suite laisse toutefois perplexe sur le niveau de robustesse aux intrusions du système en question : à partir des informations qu'il avait collectées, le jeune pirate réussit à construire un dispositif permettant la commande des aiguillage, à partir d'une simple télécommande de téléviseur ! L'accident, fait d'un jeune inconscient qui fit tout de même 12 blessés, fait réfléchir aux possibles conséquences d'une attaque de la part d'une organisation terroriste bien outillée et renseignée qui parviendrait à diffuser de fausses informations de position aux systèmes modernes de contrôle-commande ferroviaires ! La catastrophe ferroviaire œuvre d'un cyber-terroriste et non pas d'un terroriste « classique » expert en explosifs !

Rassurons toutefois le lecteur, le cas a été envisagé, et traité, pour tous les systèmes modernes récents dont font partie ceux conformes à la spécification ERTMS pour l'aspect grandes lignes, et les systèmes type CBTC pour l'urbain. Examinons toutefois dans quelle mesure les techniques de protection exposées dans la section précédente couvrent ou ne couvrent pas ce cas. Les techniques de protection d'un message par code type CRC, Hamming…, efficaces contre les fautes accidentelles sont totalement inopérantes contre les fautes malveillantes : admettant même que l'individu malveillant ignore le polynôme modulo utilisé pour un CRC par exemple, celui-ci n'a pas été conçu pour être une clé secrète d'un crypto système symétrique. Toute attaque même simple (en force brute) aurait de bonnes chances d'aboutir. Il en est de même pour un message protégé par une technique genre processeur codé avec signature : d'une part la taille de la signature sur 32 ou 48 bits est maintenant modeste par rapport à la taille recommandée actuellement pour les clés (128 voire 256 bits), mais surtout cette signature est comme on l'a souligné *statique* pour une variable donnée, et ne change donc pas d'un cycle à un autre (le contenu fonctionnel de la donnée ayant pu par contre changer). L'extraction de la signature d'une donnée et sa réinjection dans une donnée malveillante est une faute qui n'a pas été retenue (on s'en doute) comme faute naturelle d'un microprocesseur… elle est par contre à la portée d'un cyber-délinquant un peu au fait de cette technique…

A ce stade donc une conclusion s'impose : la protection contre la modification malveillante d'une information échangée est depuis l'origine du problème (probablement aussi ancienne que l'écriture) le domaine de la *cryptographie*. C'est donc bien sûr là que se situe la bonne manière de traiter le problème et pas par des techniques qui n'ont envisagé que l'accidentel… quoique comme on va le voir, il y a des parentés. Le lecteur doit tout d'abord savoir, s'il l'ignorait, que les protections cryptographiques classiques contre l'intrusion sur un réseau wifi (type clés wep) ne résistent pas aux attaques d'un cyber délinquant expérimenté. Cela va bien pour le réseau domestique (des « couches » supplémentaires permettent de garantir la confidentialité des échanges avec les sites dits « sécurisés ») qui a peu de chance d'intéresser une organisation terroriste, mais pas pour un système de communication sol / train par exemple. En revanche, il existe dans l'arsenal cryptographique une version de la protection des modifications de messages adaptée aux modifications malveillantes, c'est le domaine des *Codes d'Authentification de Messages* (Message Authentification Codes :

MAC), qui utilisent pour la plupart des fonctions de hachage combinées avec une clé secrète partagée par l'émetteur et le destinataire : on parle alors de « Keyed-Hash Message Authentification Code » (HMAC) ou code d'authentification de messages à fonction de hachage avec clé, standardisé par le National Institute of Standards and Technology [7]. Une fonction de hachage génère une empreinte de taille fixe (256 à 512 bits pour les algorithmes actuellement en usage type SHA) d'un message de taille éventuellement variable, de telle sorte qu'il est non seulement très improbable qu'une modification accidentelle d'un message génère un message qui ait la même empreinte, mais que de plus trouver (volontairement) un message ayant cette même empreinte est un problème difficile (irréalisable dans des délais raisonnables dans l'état actuel des connaissances scientifiques et des technologies informatiques). Les fonctions de hachage actuelles sont même résistantes aux collisions (il est difficile de trouver deux messages de même empreinte) pour des raisons de robustesse dans d'autres applications (dont la signature électronique mentionnée plus loin). La fonction de hachage peut donc être vue comme un code détecteur d'erreurs de haut de gamme, mais ne suffit pas à couvrir les substitutions malveillantes de message par un attaquant, car il lui suffit de substituer également l'empreinte d'origine par l'empreinte du message substitué (l'algorithme de calcul des empreintes est bien évidemment public). Il faut donc utiliser la fonction de hachage conjointement avec une clé secrète, afin de réaliser une empreinte qu'il soit impossible de falsifier sans connaître cette clé. Le code d'authentification ainsi obtenu ([7] ne n'impose pas de fonction de hachage particulière, toute fonction de hachage considérée comme robuste peut être utilisée) garantit donc l'intégrité du message ainsi que son origine de manière très similaire à la signature électronique qui pourrait d'ailleurs également être utilisée dans ce contexte (la différence étant que les algorithmes de signature électronique utilisent des crypto systèmes asymétriques ou à clé publique afin que *tout le monde* puisse vérifier l'authenticité du message). Les messages d'authentification sont donc un moyen efficace de protection contre les intrusions dans des systèmes qui doivent échanger des messages sur des media ouverts dont les communications radio, mais peuvent en définitive être vus comme un extension des codes détecteurs d'erreurs qui couvrent les modifications accidentelles de messages, aux cas des modifications à caractère malveillant.

## 5 Conclusion

Les systèmes dits « critiques » (qui peuvent présenter des dangers pour l'homme ou l'environnement) doivent être conçus pour éviter que ces dangers ne se manifestent suite à des événements que, faute de terme plus adapté, on peut appeler « accidentels » sachant que cette notion recouvre des phénomènes physiques internes ou externes (que le langage courant appelle respectivement « pannes » et « perturbations ») ainsi que des phénomènes humains sans intention de nuire en conception ou en exploitation (que le langage courant appelle respectivement « bogue » et « erreur humaine »). Tout développement de système critique intègre des moyens de couverture de ce qu'un langage plus rigoureux appelle « fautes autres qu'humaines avec intention de nuire ». La plupart de ces techniques de couverture qui permet d'obtenir le niveau de « safety » ou « sécurité innocuité » a été passé en revue au début de cet article. Mais à l'heure actuelle, un système critique se doit d'intégrer également la problématique de robustesse aux agressions humaines avec intention de nuire, en particulier pour les systèmes de plus en plus nombreux amenés à communiquer en exploitant des liaisons radio. La seconde partie de cet article a permis de passer en revue les différents types d'agression envisageables et de voir dans quelle mesure elles étaient couvertes par les techniques conçues pour la sécurité innocuité. On a vu à cette occasion le degré de parenté qu'il existait entre les deux problématiques (la couverture de la malveillance amenant souvent à une mise en œuvre avec une rigueur accrue ou avec quelques aménagements, des techniques conçues pour la sécurité innocuité). La prise en compte simultanée des deux problématiques de sécurité innocuité et sécurité confidentialité dans un concept commun de *sécurité globale* est donc certainement l'avenir pour les systèmes critiques.

## 6 Références